\documentclass[11pt,document,nofootinbib,superscriptaddress,onecolumn,preprintnumbers,balancelastpage]{article}
\pdfoutput=1
\usepackage[utf8]{inputenc}
\usepackage{jheppub}
\usepackage[english]{babel}
\usepackage{amsmath,amssymb,amsfonts}
\usepackage{graphicx}
\usepackage{booktabs}
\usepackage{hyperref}
\usepackage{xcolor}
\usepackage[sort&compress]{natbib}
\usepackage{tikz}
\usepackage{pgfplots}
\usepackage{cleveref}

\newcommand{\Neff}{N_\text{eff}}
\newcommand{\prn}[1]{ \left(  #1 \right) }
\newcommand{\al}[1]{\begin{align} #1 \end{align}}
\newcommand{\avg}[1]{\left\langle#1\right\rangle}

\begin{document}

\preprint{IPMU19-0144, DESY 19-175}
\title{Baryogenesis from a dark first-order phase transition}

\author[a,b]{Eleanor Hall,}
\author[c]{Thomas Konstandin,}
\author[a,b]{Robert McGehee,}
\author[a,d,b,c,\dagger]{Hitoshi Murayama,}
\author[c,e]{and G\'eraldine Servant}

\affiliation[a]{Department of Physics, University of California, Berkeley, CA 94720, USA}
\affiliation[b]{Ernest Orlando Lawrence Berkeley National Laboratory, Berkeley, CA 94720, USA}
\affiliation[c]{DESY, Notkestraße 85, D-22607 Hamburg, Germany}
\affiliation[d]{Kavli Institute for the Physics and Mathematics of the
  Universe (WPI), University of Tokyo,
  Kashiwa 277-8583, Japan}
\affiliation[e]{II. Institute of Theoretical Physics, University of Hamburg, D-22761 Hamburg}
\affiliation[\dagger]{Hamamatsu Professor}
\emailAdd{nellhall@berkeley.edu}
\emailAdd{thomas.konstandin@desy.de}
\emailAdd{robertmcgehee@berkeley.edu}
\emailAdd{hitoshi@berkeley.edu}
\emailAdd{geraldine.servant@desy.de}

\abstract{
We present a very minimal model for baryogenesis by a dark first-order phase transition.  It employs a new dark $SU(2)_{D}$ gauge group with two doublet Higgs bosons, two lepton doublets, and two singlets.  The singlets act as a neutrino portal that transfer the generated asymmetry to the Standard Model.  The model predicts $\Delta \Neff = 0.09$--0.13 detectable by future experiments as well as possible signals from exotic decays of the Higgs and $Z$ bosons and stochastic gravitational waves.
}
\maketitle

\begin{center}
{\it This paper is dedicated to the memory of Ann Elizabeth Nelson.}
\end{center}

\section{Introduction}

The origin of the baryon asymmetry of the universe (BAU) remains an open question. Although baryon number is conserved at tree level by the Standard Model (SM), the present-day matter density suggests an asymmetry between baryons and anti-baryons in the early universe at the level of one part in a billion.
Resolving this BAU question has become more urgent with the recent success of inflation in high-precision tests of the anisotropy of the cosmic microwave background (CMB) \cite{Akrami:2018odb}. Even if some baryon asymmetry existed at the beginning of the Universe, inflation would have diluted it by $e^{-N}$ where the $e$-fold $N$ needs to be larger than 50 in order to solve the horizon and flatness problems~\cite{Guth:1980zm}.  Therefore, the present-day BAU must have been generated after inflation by a micro-causal mechanism. Such a mechanism must satisfy three conditions, as pointed out by Sakharov~\cite{Sakharov:1967dj}: (1) violation of baryon number, (2) violation of C and CP, and (3) departure from thermal equilibrium.  

While there are many possible mechanisms for creating a baryon asymmetry, there are two general directions that are popular in the literature. One is {\it leptogenesis} \cite{Fukugita:1986hr}, which is an automatic consequence of the origin of the small neutrino mass from the so-called seesaw mechanism \cite{Yanagida:1980xy,Minkowski:1977sc,GellMann:1980vs}. Unfortunately, this mechanism is difficult to test experimentally because it relies on physics at very high-energy scales (see, however Ref.~\cite{Chun:2017spz} and a recent discussion on a potential test using gravitational waves in Ref.~\cite{Dror:2019syi}).

The other popular mechanism which explains the BAU is {\it electroweak baryogenesis} (see Ref.~\cite{Cohen:1993nk} by Andrew Cohen, David B. Kaplan, and Ann E. Nelson for a pioneering review and \cite{Morrissey:2012db,Konstandin:2013caa} for recent updates). 
This mechanism is motivated by the fact that the SM in principle satisfies all of the Sakharov conditions, with B violation in the anomalous $SU(2)$ sphaleron process, C violation in the weak interaction, CP violation in the CKM matrix, and departure from thermal equilibrium in a first-order electroweak phase transition. However, the degree of CP violation in the CKM matrix is too small to account for the needed asymmetry \cite{Jarlskog:1985ht,Gavela:1994dt, Huet:1994jb}, and with the observed Higgs boson mass of 125~GeV, the SM electroweak phase transition is crossover \cite{Kajantie:1995kf,Kajantie:1996qd,Rummukainen:1998as}. Mechanisms for electroweak (EW) baryogenesis therefore must introduce some additional particle content such as singlet scalars \cite{Espinosa:2011eu} and extended higgs sectors \cite{Fromme:2006cm} in order to fully realize conditions (2) and (3). This new content is often accessible at high-energy colliders such as the Large Hadron Collider (LHC) and may be tested by precision measurements at much lower energies, making it very falsifiable.

Unfortunately these models are, in some sense, too falsifiable.  They tend to predict an electric dipole moment (EDM) of the electron, neutron, and atoms at levels which are highly constrained by recent experimental results \cite{Andreev:2018ayy}. Thus, it is worthwhile to look for theories that achieve EW baryogenesis or something similar and can be tested by current or future experiments. Recent attempts in this direction have considered CP violation from strong CP violation \cite{Servant:2014bla}, varying Yukawas \cite{Servant:2018xcs}, SM leptons \cite{deVries:2018tgs,Bruggisser:2017lhc}, a dark sector \cite{Cline:2017qpe,Carena:2018cjh}, and higher-scale sources if the EW phase transition happens at higher temperatures \cite{Baldes:2018nel,Glioti:2018roy}.

In this paper, we propose a model that achieves baryogenesis at energies just above the EW scale with very few new degrees of freedom through electroweak-like baryogenesis in a dark sector. In our model, a first-order phase transition in a dark sector with two Higgs doublets generates an asymmetry through the charge transport mechanism, with ``baryon'' number violation from an anomalous dark gauge group and CP violation from a non-trivial phase in the dark Higgs. The dark sector is connected to the SM by a renormalizable neutrino portal, and so the dark-sector asymmetry is converted into a SM baryon asymmetry through the SM sphaleron. Our model closely resembles ``darkogenesis'' by Shelton and Zurek \cite{Shelton:2010ta}. Unlike darkogenesis, our model uses the neutrino portal instead of a messenger sector or higher-dimensional operator and thus is fully renormalizeable; we do not attempt to realise asymmetric dark matter.

Producing a primordial asymmetry in a dark sector which is then transferred to the visible sector without requiring violation of baryon or lepton number beyond the SM (BSM) was previously explored in \cite{Servant:2013uwa} (case II).
There, the origin of the dark asymmetry was not specified; the emphasis was on using a higher-dimensional Higgs portal for the transfer and realizing asymmetric dark matter.
In this paper, we provide a UV completion in which the transfer operator involves new singlet leptons and the Higgs such that the EW phase transition does not have to be first-order, only the dark phase transition has to be.
In addition, our model predicts new relativistic degrees of freedom in the Universe today at the level detectable by near-future CMB experiments.  Furthermore, it retains the salient feature of EW baryogenesis that leads to the stochastic gravitational wave signature. 

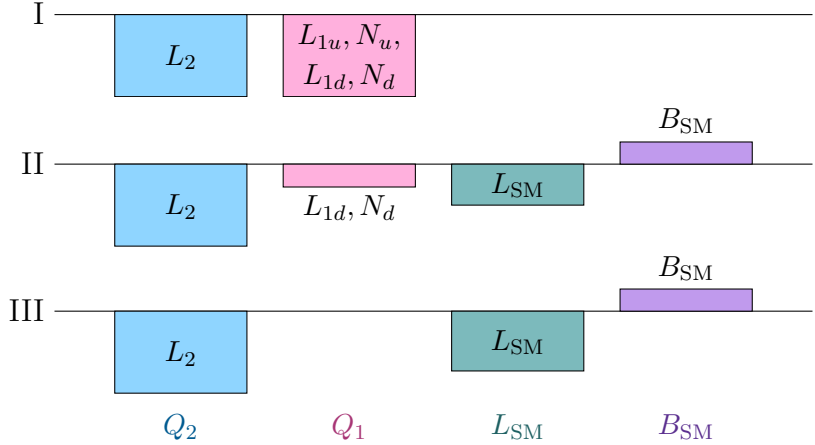
\begin{figure}
\centering
\begin{tikzpicture}
  \definecolor{blue}{HTML}{8fd5ff}
  \definecolor{pink}{HTML}{ffb0dd}
  \definecolor{green}{HTML}{7CBABC}
  \definecolor{purple}{HTML}{C09AED}

  \definecolor{darkblue}{HTML}{005b91}
  \definecolor{darkpink}{HTML}{a83676}
  \definecolor{darkgreen}{HTML}{1f6163}
  \definecolor{darkpurple}{HTML}{5b2f8f}
  \pgfplotsset{
    axis y line=none,
    xtick=\empty,
    axis lines*=center,
    nodes near coords,
    point meta=explicit symbolic,
    width = 0.75\textwidth,
    ymin = -1,
    ymax = 0.3,
    xmin = 1,
    xmax = 4,
    height=3cm,
    enlarge x limits=0.25,
    xlabel style={at={(0,1.15)},left},
    visualization depends on=(-8*(y<0)*(y>-0.3)+ (15*abs(y))*(y<-0.3) + 8*(y>0)) \as \myshift,
    every node near coord/.append style={align=center,yshift=\myshift,anchor=center},
    every axis plot/.append style={ybar,bar width=50, symbolic x coords={1,2,3,4}, bar shift=0pt}
  }
  \begin{axis}[name=plot1,xlabel={\large I}]
    \addplot[fill=blue] coordinates {(1,-1)[{\color{black}$L_2$}]};
    \addplot[fill=pink] coordinates {(2,-1)[{\color{black} $L_{1 u}, N_u,$ \\ \color{black}$L_{1 d},N_d$}]};
  \end{axis}
  \begin{axis}[name=plot2,xlabel={\large II}, at=(plot1.below south west), anchor=above north west]
    \addplot[fill=blue] coordinates {(1,-1)[{\color{black}$L_2$}]};
    \addplot[fill=pink] coordinates {(2,-0.279)[{\color{black}$L_{1 d}, N_d$}]};
    \addplot[fill=green] coordinates {(3,-0.5)[{\color{black}$L_\text{SM}$}]};
    \addplot[fill=purple] coordinates {(4, 0.27)[{\color{black}$B_\text{SM}$}]};
  \end{axis}
  \begin{axis}[name=plot3,xlabel={\large III}, at=(plot2.below south west), anchor=above north west]
    \addplot[fill=blue] coordinates {(1,-1)[{\color{black}$L_2$}]};
    \addplot[fill=green] coordinates {(3,-0.729)[{\color{black}$L_\text{SM}$}]};
    \addplot[fill=purple] coordinates {(4, 0.27)[{\color{black}$B_\text{SM}$}]};
  \end{axis}
  \begin{axis}[name=plot4, at=(plot3.below south west), anchor=above north west,
    height=1.6cm,
    axis x line*=bottom,
    axis line style = { draw = none },
    xtick style={draw=none},
    xtick={1,2,3,4},
    xticklabels={{\color{darkblue}$Q_2$},{\color{darkpink}$Q_1$},{\color{darkgreen} $L_\text{SM}$},{\color{darkpurple} $B_\text{SM}$}}]

  \end{axis}
\end{tikzpicture}
\caption{The schematics of the evolution of asymmetries. During step I, reflection by bubble walls and the dark SU(2) sphaleron generate the initial dark sector asymmetries. In step II, $N_u$ decays to $L_{SM}$ and the SM sphaleron actively converts some of this asymmetry to $B_{SM}$. $N_d$ decays to $L_{SM}$ in step III.}
\label{fig:scheme}
\end{figure}

\section{Basic Idea}
We employ an $SU(2)_{D}$ gauge theory with two Higgs doublets.  Here, $D$ stands for ``dark'' and we refer to the equivalent SM gauge group as $SU(2)_{\rm SM}$.  We introduce one set of ``leptons'' that consists of a left-handed $SU(2)_{D}$ doublet $L_{1} = (L_{1u}, L_{1d})$, and two right-handed singlets $N_{u,d}$ (note that they do not form a doublet under $SU(2)_D$).   One eigenstate ``top lepton'' has $O(1)$ Yukawa couplings and plays the role of the top quark in the original EW baryogenesis, while the other eigenstate ``bottom lepton'' is analogous to the bottom quark, and we assume it has a much smaller Yukawa coupling so that we can ignore it from the dynamics of the bubble walls.  In order to cancel Witten’s anomaly, we need another doublet $L_2$, but we do not introduce accompanying right-handed fermions to prevent any leakage of $L_2$ charge into SM $B+L$.   
We also impose a ${\mathbb Z}_{2}$ symmetry under which $L_{2}$ is the only odd field to forbid the mass term $L_{1} L_{2}$.

We assume the phase transition is first order and happens before EW symmetry breaking.  The potential of the two Higgs doublets is
\begin{equation}
\begin{split}
	V(\Phi) &=  \mu_{1}^{2} \Phi_{1}^{\dagger} \Phi_{1} + \mu_{2}^{2} \Phi_{2}^{\dagger} \Phi_{2}
	- \mu_{3}^{2} (\Phi_{1} ^{\dagger} \Phi_{2} + c.c.) \\
	& + \frac{1}{2} \lambda_{1} (\Phi_{1}^{\dagger} \Phi_{1})^{2}
	+ \frac{1}{2} \lambda_{2} (\Phi_{2}^{\dagger} \Phi_{2})^{2}  
	+ \lambda_{3} (\Phi_{1}^{\dagger} \Phi_{1}) (\Phi_{2}^{\dagger} \Phi_{2})
	+ \lambda_{4} (\Phi_{1}^{\dagger} \Phi_{2}) (\Phi_{2}^{\dagger} \Phi_{1}) \\
	& +\left[ \frac{1}{2} \lambda_{5} (\Phi_{1}^{\dagger} \Phi_{2})^{2}
	+ \lambda_{6} (\Phi_{1}^{\dagger}\Phi_{1}) (\Phi_{1}^{\dagger} \Phi_{2})
	+ \lambda_{7} (\Phi_{1}^{\dagger}\Phi_{2}) (\Phi_{2}^{\dagger} \Phi_{2})
	+c.c. \right].
\end{split}
\end{equation}
The couplings $\lambda_{5,6,7}$ are complex and their imaginary parts violate CP.  
The Yukawa couplings consistent with the ${\mathbb Z}_{2}$ symmetry are
\begin{align}
	{\cal L}_{Y} = & -Y_{a\alpha} \bar{L}_{1} \Phi_{a} N_{\alpha} 
	-\tilde{Y}_{a\alpha} \bar{L}_{1} \tilde{\Phi}_{a} N_{\alpha} 
	+ c.c.
\end{align}
Here, $\tilde{\Phi}_a = i \sigma_{2} \Phi^{*}_a$, and $a=1,2$, $\alpha=u,d$.  $L_{2}$ remains exactly massless while $N_\alpha$ carry $Q_1$ charge.  At this stage, we find an exact $Q_{1}-Q_{2}$ symmetry.  

\begin{table}[t]
\centering
\begin{tabular}{|c|c|c|c|c|c|c|}
\hline
field & $SU(2)_D$ & $\gamma_{5}$ & $Q_{1}$ & $Q_{2}$ & ${\mathbb Z}_{2}$ \\ \hline
$\Phi_{1,2}$ & {\bf 2} & 0 & 0 & 0 & $+$\\ \hline
$L_{1}$ & {\bf 2} & $-1$ & $+1$ & 0 & $+$\\ \hline
$N_{u,d}$ & {\bf 1} & $+1$ & $+1$ & 0 & $+$\\ \hline
$L_{2}$ & {\bf 2} & $-1$ & $0$ & $+1$ & $-$\\ \hline
\end{tabular}
\caption{The particle content of the dark sector. }
\end{table}

The leptons $L_{1}$ and $N_{u,d}$ play the role of the top quark to produce the $Q_{1}+Q_{2}$ asymmetry.  Since $Q_{1}-Q_{2}$ is conserved by the $SU(2)_{D}$ sphaleron, the generated asymmetries satisfy $Q_{1} = Q_{2}$.  This is the first step in Fig.~\ref{fig:scheme}.
On the other hand, the $Q_{1}$ charge can equilibrate with the Standard Model leptons $\ell_{i}$ through the Yukawa couplings
\begin{equation}
	\Delta {\cal L}_{Y}
	= -y_{i \alpha} \bar{\ell}_{i} N_{\alpha} \tilde{H} + c.c.,
\end{equation}
where $H$ is the standard model Higgs doublet and $\tilde{H} = i \sigma_{2} H^{*}$. The conserved (non-anomalous) quantity is then
\begin{equation}
	Q \equiv Q_{1} - Q_{2} + L_{\rm SM} - B_{\rm SM}.
\end{equation}
As the ``top lepton" decays into the SM leptons, the lepton asymmetry $L_{\rm SM}$ is then generated, which is partially converted to the baryon asymmetry $B_{\rm SM}$ through the sphaleron transitions in $SU(2)_{\rm SM}$.  This is the second step in Fig.~\ref{fig:scheme}. Finally, the SM sphaleron freezes out and $B_{\rm SM}$ becomes fixed, while the ``bottom lepton" decays and the remaining $Q_1$ is converted to $L_{\rm SM}$.  This is the last step in Fig.~\ref{fig:scheme}.  Note that it is also possible for the ``bottom lepton" to decay into the SM before the sphaleron freeze-out, depending on its mass.

The most general Lagrangian consistent with the symmetries includes also Majorana masses for $N_{u,d}$,
\begin{equation}
	{\cal L}_{M} = \frac{1}{2} m_{\alpha\beta} N_{\alpha} N_{\beta} 
	+ c.c. 
	\label{eq:Majorana}
\end{equation}
This term violates $Q$.  In order to maintain the baryon asymmetry, we need to make sure that $Q$ violation is small.  Approximately, we need
\begin{equation}
	\frac{m^{2}}{T_*^{2}} < \frac{T_*}{M_{Pl}},
\end{equation}
at the time of the sphaleron freeze-out at $T_{*}=(131.7\pm 2.3)$~GeV \cite{DOnofrio:2014rug}, and hence $m \lesssim {\rm keV}$.  With or without the Majorana mass terms, there are three massless states (see Eq.~\eqref{eq:massmatrix} below where the mass matrix is rank 2).  They make massive fermions pseudo-Dirac, namely split Dirac fermions into nearly degenerate two Majorana fermions each. The light neutrino masses as observed by neutrino oscillation must come from another source, such as the popular seesaw mechanism at high energy scales.  Therefore, the Majorana mass terms in Eq.~\eqref{eq:Majorana} are unimportant for phenomenology as long as the bound is satisfied, and we will ignore them in the discussions below.

Once all Higgs fields acquire expectation values $\langle H \rangle = v$ and $|\langle \Phi_1 \rangle|^2 + |\langle \Phi_2 \rangle|^2 = V^2$,
the neutral lepton sector has a mass matrix
\begin{align}
	\left(
	\begin{array}{ccccc}
		\bar{L}_{1u} & \bar{L}_{1d} & \bar{\nu}_{e} & \bar{\nu}_{\mu} & \bar{\nu}_{\tau}
	\end{array} \right)
	\left( \begin{array}{cc}
		M_{u} & 0 \\
		0 & M_{d} \\
		y_{e1} v & y_{e2} v \\
		y_{\mu 1} v & y_{\mu 2} v \\
		y_{\tau 1} v & y_{\tau 2} v
	\end{array} \right)
	\left( \begin{array}{c} N_{u} \\ N_{d} \end{array} \right). 
	\label{eq:massmatrix}
\end{align}
Here we have made an $SU(2)_{D}$ gauge rotation as well as the $U(2)$ basis rotation of $N_{u,d}$ to diagonalize the upper $2\times 2$ block to ${\rm diag}(M_{u}, M_{d})$ where both eigenvalues $M_{\alpha}$ are real and positive.  Henceforth we refer to the dark lepton states in this basis where $L_{1u}, N_{u}$ refer to the top lepton, while $L_{1d}, N_{d}$ to the bottom lepton.
Then the massive Dirac eigenstates are given approximately by
\begin{align}
L'_{1u} &= L_{1u} + \epsilon_{iu} \nu_{i}, \\
L'_{1d} &= L_{1d} + \epsilon_{id} \nu_{i}, 
\end{align}
for the top and bottom leptons, respectively, while the massless states are given by
\begin{equation}
\nu'_{i} = \nu_{i} - \epsilon_{iu}^{*} L_{1u} - \epsilon_{id}^{*} L_{1d} .
\end{equation}
Here, $\epsilon_{i\alpha} = y_{i\alpha} v / M_{\alpha}$.
For the discussions below, we assume $V \sim $1--100~TeV, and $M_{u} \approx V$, but it is easy to see how phenomenology changes for different parameters.

\section{Asymmetries}

We assume that the dark sector undergoes a first-order phase transition.  The phases of the Higgs fields vary inside the bubble walls due to the CP-violating couplings in the potential.  This CP violation affects the reflection coefficients of the (dark) top lepton, which induces the asymmetry in $Q_{1}$ and also $Q_{2}$.  

Since the sphaleron in $SU(2)_{D}$ preserves $Q_{1}-Q_{2}$, we find
\begin{equation}
	Q_{1} = Q_{2} \neq 0.
\end{equation}
If the dark phase transition is strong enough, the dark sphaleron will be suppressed subsequently and washout is avoided.  From this point on, there are no interactions that can change $Q_2$, and hence the asymmetry is stored and protected.  
The $Q_{1}$ charge transforms to Standard Model leptons by the decay $N_u \rightarrow \nu_{i} h, \nu_i m_Z$ with the rate
\begin{align}
    \Gamma(N_u) = \frac{1}{32\pi} |\epsilon_{iu}|^2 \frac{M_u^3}{v^2} \beta_f(M_u,m_h)^2
    &+ \frac{1}{64\pi} g_Z^2 |\epsilon_{iu}|^2 M_u \left(2 + \frac{M_u^2}{m_Z^2}\right) \beta_f(M_u,m_Z)^2,
\end{align}
where the phase space factor is
\begin{equation}
	\beta_f(M,m) = 1 - \frac{m^{2}}{M^{2}}\ . 
	\label{eq:beta}
\end{equation}
For this process to reach equilibrium before the sphaleron process freezes out, we need $\Gamma(N_u)>H(T_{*})$.  Assuming $M_u \gg m_h, m_Z$,
\begin{equation}
	|\epsilon_{iu}|^2 > 5.02 \times 10^{-17}
	\times \left( \frac{T_{*}}{131.7~\mbox{GeV}} \right)^2
	\left( \frac{\rm TeV}{M_u} \right)^3
	\ ,
\end{equation}
a very weak constraint.  Hereafter we assume it is satisfied.

Most SM interactions are in equilibrium at this stage as is the SM weak sphaleron, but the charge combination 
$L_{SM} + Q_1 - B_{SM}$ is conserved. 
Using the established approach \cite{Harvey:1990qw}, the initial $Q_1$ spreads across the SM degrees of freedom. The actual numbers then depend on when the neutrinos that carry the $Q_1$ charge decay and whether the EW phase transition in the SM is strongly first order or not. We always assumed $N_u$ decays before EW symmetry breaking so far, but now consider the other case as well.

We first consider the schematics in Fig.~\ref{fig:scheme} where $N_d$ decays after the SM sphalerons become inefficient.  This is the most interesting scenario because it provides the collider signatures discussed in the next section.  Without any additional new particles, the SM phase transition is a crossover, where sphaleron effects continue to exist down to $T_*$.  Then the chemical equilibrium is achieved in the broken phase, and one finds 
\begin{equation}
	B_{\rm SM} =- \frac{36}{133} Q_{2} \, , \quad 
	L_{\rm SM} = \frac{97}{133} Q_{2} \, .
\end{equation}
If the SM phase transition is strongly first order instead, the sphaleron freezes out immediately after the phase transition.  Then the chemical equilibrium achieved in the unbroken phase determines the asymmetries, yielding
\begin{equation}
	B_{\rm SM} = -\frac{28}{101} Q_{2} \, , \quad 
	L_{\rm SM} = \frac{73}{101} Q_{2} \, .
\end{equation}
If $N_d$ is heavy, it may decay before the EW sphalerons freeze out. This scenario yields 
\begin{equation}
	B_{\rm SM} = -\frac{12}{37} Q_{2} \, , \quad 
	L_{\rm SM} = \frac{25}{37} Q_{2} \, ,
	\label{eq:HTcrossover}
\end{equation}
when the SM is crossover, or 
\begin{equation}
	B_{\rm SM} = -\frac{28}{79} Q_{2} \, , \quad 
	L_{\rm SM} = \frac{51}{79} Q_{2} \, .
\end{equation}
if the SM has a strongly first-order phase transition.  These are the same results as in  Ref.~\cite{Harvey:1990qw}. 

The final question is if sufficient initial $Q_1$ charge can be produced from the $SU(2)_D$ phase transition. The picture we have in mind is 
akin to the usual EW baryogenesis in the two Higgs doublet model (see \cite{Fromme:2006cm}).
However, there are several factors that work in favor of the dark sector. First, since the spectrum of the dark scalars is not constrained, the phase transition does not need to rely on the interplay of the two scalars. If the Higgses are light enough, strong enough phase transitions can be obtained through the thermal contributions of the $SU(2)_D$ gauge bosons, $\left<\Phi \right>/T \simeq g_D^3/\lambda$, where $\lambda$ is a generic quartic scalar coupling in the dark sector (notice that this also solves issues with Landau poles that often occur in the two-Higgs doublet extensions of the SM setup). Second, all dark Higgses can be of similar mass which tends to increase the change of the complex phases in the Higgs fields during the phase transitions. Third, the $SU(2)_D$ gauge coupling could be substantially larger than that of $SU(2)_{\rm SM}$ and hence enhance the BAU due to a larger (dark) sphaleron rate and a stronger phase transition. Finally, the $L_1$ fields do not carry color and hence diffuse farther into the symmetric phase and do not suffer from suppression by the strong sphalerons. Altogether, we expect that the model can potentially produce a BAU that is a few orders of magnitude larger than the observed one. 

\section{Laboratory Signatures}

Lepton universality in $\tau^{-} \rightarrow \mu^{-} \bar\nu_{\mu} \nu_{\tau}$, $\tau^{-} \rightarrow e^{-} \bar\nu_{e} \nu_{\tau}$, and $\mu^{-} \rightarrow e^{-} \bar\nu_{e} \nu_{\mu}$ is tested at the permille level~\cite{Pich:2013lsa} which implies $1-|\epsilon_{iu}|^{2} -|\epsilon_{id}|^{2}$ are the same among $i=e,\mu,\tau$ at the level of $10^{-3}$. Barring the conspiracy where $\epsilon_{i\alpha}$ are the same to all three $i$, we typically need $|\epsilon_{i\alpha}|^{2} \lesssim 10^{-3}$.  Improved measurements of $\tau$ properties at Belle II may be able to discover non-universality.

If $N_{d}$ is lighter than $Z$, the decay $Z \rightarrow N_{d} \bar\nu_{i} + \bar{N}_{d} \nu_{i}$ has the branching fraction
\begin{align}
	{\rm BR}(Z \rightarrow N_{d} \bar\nu_{i} + \bar{N}_{d} \nu_{i})
	&= |\epsilon_{id}|^{2} {\rm BR}(Z \rightarrow \nu_i \bar{\nu}_i) \beta_f^2 (3-\beta_f)
	\nonumber \\
	&= 0.067 |\epsilon_{id}|^{2} \beta_f^2 (3-\beta_f).
\end{align}
$N_{d}$ subsequently decays as $N_{d} \rightarrow \ell_{j}^{-} q\bar{q}', \ell_{j}^{-} \ell_{k}^{+} \nu_{k}$ picking up the mixing $\epsilon_{jd}$. 
The search for neutral heavy leptons was performed by DELPHI at LEP~\cite{Abreu:1996pa}, and the upper limit on the mixing angle squared is as strong as $|\epsilon_{id}|^{2} < 2\times 10^{-5}$ for a range of masses and decay lengths, while is weaker for $M_d \gtrsim 50$~GeV and reverts to the limit from universality once $M_d > m_Z$.  For very light $M_d \lesssim 2$~GeV, there are stronger limits from fixed-target experiments. Future $Z$ factories (GigaZ at ILC or TeraZ at FCC$ee$) will better probe this decay.

If $N_{d}$ is lighter than the Higgs boson of mass $m_{h} = 125$~GeV, the Higgs boson can decay into $\bar{N}_{d} \nu_{i} + N_{d} \bar\nu_{i}$.  
Using BR$(h \rightarrow \tau^{+} \tau^{-}) = 6.3 \times 10^{-2}$ \cite{Tanabashi:2018oca}, we find
\begin{equation}
\begin{split}
	{\rm BR}(h \rightarrow \bar{N}_{d} \nu_{i} + N_{d} \bar\nu_{i})
	&= 6.3 \times 10^{-2}
	\times \left| \frac{\epsilon_{id} M_{d}}{m_{\tau}} \right|^{2} \beta_f^2
	\\
	&= 1.80\times 10^{-4} \frac{|\epsilon_{id}|^{2}}{10^{-5}} 
	\left( \frac{M_{d}}{30~{\rm GeV}} \right)^{2} \beta_f(m_h,M_d)^2.
\end{split} 
\end{equation}
This can be sizable and appears as an exotic decay of the Higgs boson.  This can be probed down to the level of $10^{-4}$ or better at future $e^{+}e^{-}$ Higgs factories \cite{Liu:2016zki}. 

There is no contribution to the electric dipole moment of quarks in this model.  That of the electron is suppressed by $|\epsilon_{eu}|^{2}$, $|\epsilon_{ed}|^{2}$.  This suppression factor makes the proposed model of baryogenesis here perfectly compatible with the stringent constraint from the ACME collaboration \cite{Andreev:2018ayy}.

\section{Excess radiation and $\Delta \Neff$}

$L_{2}$ has a small asymmetry, but since it is massless, it has a thermal abundance and can contribute non-negligibly to the energy density of the early, radiation-dominated Universe. The total energy density, $\rho_r$, is parameterized by its relation to the photon energy density, $\rho_\gamma$, via
\al{
\rho_r = \prn{1+\frac{7}{8} \prn{\frac{4}{11}}^{4/3} \Neff} \rho_\gamma,
}
where $\Neff$ is the effective number of neutrinos. The SM neutrinos contribute 3.046~\cite{Mangano:2001iu,deSalas:2016ztq} to $\Neff$ (due to their incomplete decoupling by the time of electron-positron annihilation), and in general, any relativistic BSM particles contribute as well. \emph{Planck} recently measured $\Neff = 2.99^{+0.34}_{-0.33}$ (95\% CL) \cite{Abazajian:2019eic}, while the measurements of primordial abundances from Big bang nucleosynthesis (BBN) imply $\Neff = 2.85\pm0.28$~\cite{Cyburt:2015mya}. Both of these measurements are consistent with the SM prediction and constrain any BSM relativistic species.

\begin{figure*}[t]
\centering
\includegraphics[width=\textwidth]{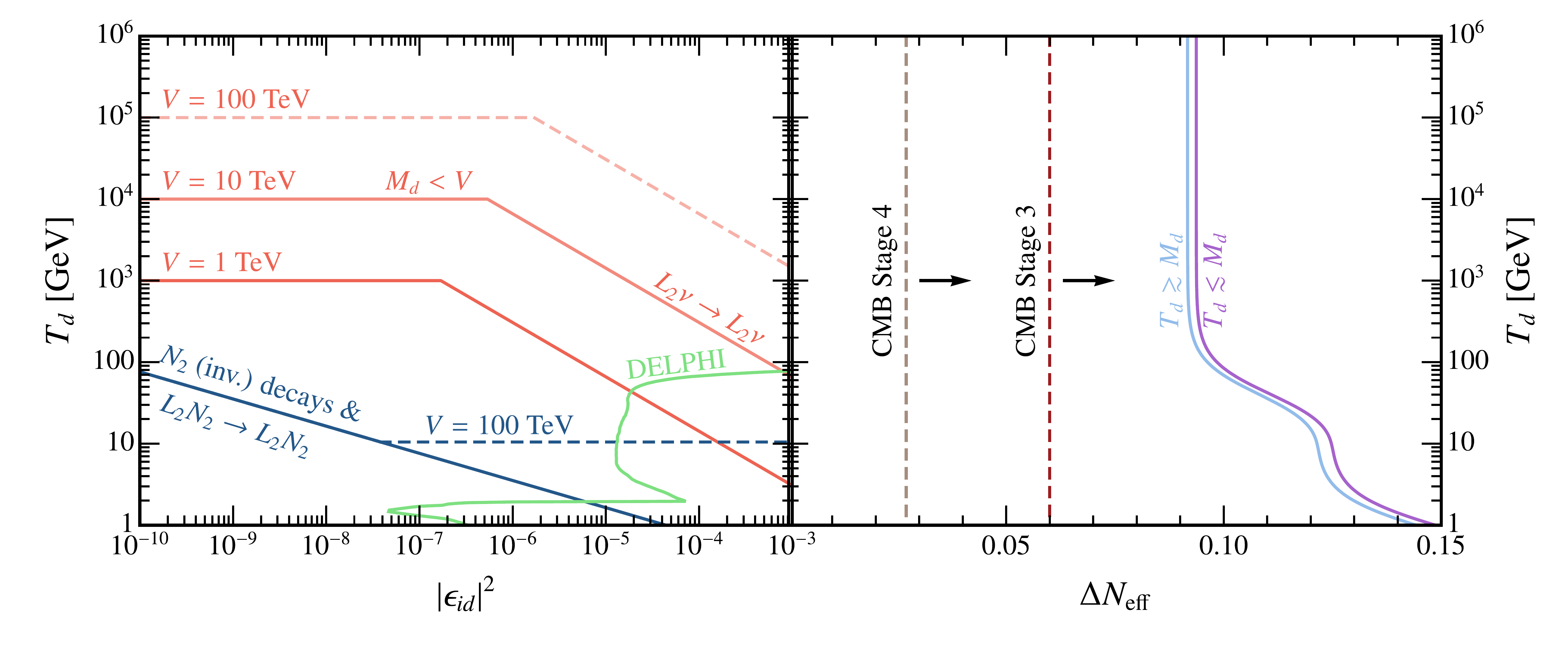}
\caption{\textbf{Left:} The schematic plot of the decoupling temperature 
$T_d$ as a function of the mixing $|\epsilon_{id}|^2$.
The red (blue) curves act as upper (lower) limits on the decoupling temperature. In case $M_d$ is between these two bounds, the decoupling temperature is given by $M_d$. See text for details.
The DELPHI limit \cite{Abreu:1996pa} requires $|\epsilon_{id}|^2$ to be on the left of the green curve.  When green and red lines cross, the red line below the green curve is still possible if $|\epsilon_{iu}|^2$ replaces the role of $|\epsilon_{id}|^2$. \textbf{Right:} $\Delta N_{\rm eff}$ for various decoupling temperatures $T_d$ of $L_2$. Also shown are the expected sensitivities at CMB Stage 3 and 4 experiments.}
\label{fig:TdandNeff}
\end{figure*}

To determine $L_2$'s contribution to $\Neff$, it is necessary to track when it kinetically decouples from the SM bath. 
$L_2$ can equilibrate with the SM neutrinos via $W_{D}^3$ exchange picking up the small $L_{1}$ component in $\nu'_{i}$ with the cross section
\begin{align}
	\sigma (L_{2} \nu'_{i} \rightarrow L_{2} \nu'_{i})
	&= \frac{1}{16\pi} (|\epsilon_{iu}|^{2}-|\epsilon_{id}|^{2})^2 \frac{s}{V^{4}} \ , \nonumber \\
	\sigma (L_{2} \bar{\nu}'_{i} \rightarrow L_{2} \bar{\nu}'_{i})
	&= \frac{1}{48\pi} (|\epsilon_{iu}|^{2}-|\epsilon_{id}|^{2})^2 \frac{s}{V^{4}} \ ,
    \label{eq:sigmaL2nu}
\end{align}
where $s$ is the usual Mandelstam variable.
The thermal average of $s$ yields
\begin{align}
    \langle s \rangle
    = 2 \left( \frac{\rho}{n} \right)^2
    = 2 \left( \frac{7\pi^4}{180 \xi (3)} \right)^{2} T^2.
\end{align}  
Assuming the mixing angles are dominated by one neutrino flavor $\nu_i$, the rate for scattering is
\begin{align}
\Gamma = n_{\nu'_i} \avg{\sigma v}(L_{2} \nu'_{i} \rightarrow L_{2} \nu'_{i})
+ n_{\bar\nu'_i} \avg{\sigma v}(L_{2} \bar\nu'_{i} \rightarrow L_{2} \bar\nu'_{i}) \, ,
\end{align}
where $n_{\nu'_i}$ is the number density of the SM neutrino. 
When this rate drops below the Hubble rate,
\al{
H = \sqrt{\frac{g_\ast \pi^2}{90}} \frac{T^2}{M_{\text{Pl}}}
}
$L_2$ falls out of equilibrium.
This occurs at
\begin{equation}
    T = 66.0~{\rm GeV} \left( \frac{g_*(T_d)}{103.9} \right)^{1/6}
    \left| \frac{10^{-5}}{|\epsilon_{iu}|^{2}-|\epsilon_{id}|^{2}} \right|^{2/3}
    \left( \frac{V}{\rm TeV} \right)^{4/3}.
    \label{eq:TdL2nu}
\end{equation}
This is the declining upper red line in Fig.~\ref{fig:TdandNeff}. 

It is also possible that $N_{d}$ acts as a mediator in equilibrating $L_2$. This requires that $N_d$ is light enough to be abundant ($T\gtrsim M_d$) and also that the interactions  of $N_d$ are strong enough to maintain equilibrium both with $L_2$ and the SM particles. The equilibrium between $L_2$ and $N_d$ is due to cross sections 
\begin{align}
	\sigma (L_{2} L_{1d} \rightarrow L_{2} L_{1d})
	&= \frac{1}{16\pi} \frac{s}{V^{4}} \ , \nonumber \\
	\sigma (L_{2} \bar{L}_{1d} \rightarrow L_{2} \bar{L}_{1d})
	&= \frac{1}{48\pi} \frac{s}{V^{4}} \ ,
\end{align}
and persists down to the temperature
\begin{equation}
    T >  22.6~\mbox{MeV} \left( \frac{g_*}{16.89} \right)^{1/6}
    \left( \frac{V}{\rm TeV} \right)^{4/3}.
    \label{eq:L1L2scatter}
\end{equation}
Except for the highest values of $V$ ({\it e.g.}\/, $V=100$~TeV shown as the dashed blue line in Fig.~\ref{fig:TdandNeff}), this process is in equilibrium for $T\gtrsim M_d$.

As for the equilibrium between $N_d$ and SM, first study the case when $M_d<m_Z$.  The equilibrium with the SM is due to its decay and inverse decay through an off-shell $Z$-boson exchange with the rate
\begin{equation}
    \Gamma (N_{d} \rightarrow \nu_i f \bar{f}) 
    = \frac{N}{768\pi^3} | \epsilon_{id}|^2 G_F^2 M_d^5,
\end{equation}
where $N$ is the effective number of neutrinos in the final state.  Actually, for $f=\nu_i$, the Fermi statistics doubles this contribution.  By adding also $f=e,\mu$ contributions, $2[(2s_W^2)^2 + (1-2s_W^2)^2]=1.01$, we have $N=5.01$.  For $M_d$ higher than $\sim m_{\pi^0}$, additional contributions from hadrons need to be included.  The decay and inverse decay are then in equilibrium down to
\begin{equation}
    T > 1.64~\mbox{GeV} \left( \frac{5.01}{N} \right)^{1/3}
    \left( \frac{g_*(T)}{86.31} \right)^{1/6} 
    \left( \frac{10^{-5}}{|\epsilon_{id}|^2} \right)^{1/3}.
    \label{eq:N2decay}
\end{equation}
This is shown as the declining blue line in Fig.~\ref{fig:TdandNeff}.  

When $M_d > m_Z, m_h$, the decay and inverse decay $N_d \leftrightarrow \nu_i Z, \nu_i h$ is in equilibrium by $T=M_d$  
as long as $|\epsilon_{id}|^2 \gtrsim 10^{-14} (M_d/{\rm TeV})^{-1}$.  We do not consider such small mixing angles below.  Therefore, once $M_d>m_Z,m_h$, the equilibrium is established before 
$T = M_d$.  On the other hand, we expect $M_d < V$ from the perturbativity which cuts off the allowed region at $V$. 

In summary, $L_2$ can stay in equilibrium either through direct interactions with the SM down to temperatures as in (\ref{eq:TdL2nu}) or through $N_d$ if (\ref{eq:L1L2scatter}), (\ref{eq:N2decay}) and $T\gtrsim M_d$ are met. In Fig.~\ref{fig:TdandNeff} the red (blue) curves act as upper (lower) limits on the decoupling temperature. In case $M_d$ is between these two bounds, the decoupling temperature is given by $M_d$.
Recall that $M_d$ can only be below the green curve from the DELPHI constraints if $|\epsilon_{iu}|^2$ replaces the role of $|\epsilon_{id}|^2$.

Once $L_{2}$ (a doublet under $SU(2)_{D}$) kinetically decouples from the SM bath, its 
energy density is subsequently diluted relative to the SM bath. After the SM neutrinos decouple from the bath, the remaining energy density in $L_{2}$ adds to $\Neff$:
\begin{equation}
	\Delta \Neff = 2 \left( \frac{g_{*s}^{\nu}}{g_{*s}^{\text{dec}}} \right)^{4/3}
\end{equation}
where $g_{*s}^{\text{dec}}$ is the number of relativistic degrees of freedom in the SM bath just after $L_{2}$ kinetically decouples and
\begin{equation}
	g_{*s}^{\nu} = 2(\gamma) + \frac{7}{8} (3\times 2(\nu)
	+ 4\times 2(e)) = 10.75
\end{equation}
is the number of relativistic degrees of freedom when the SM neutrinos decouple just before BBN. 

We show $\Delta \Neff$ in Fig.~\ref{fig:TdandNeff} as a function of the decoupling temperature $T_d$ for $L_2$, for two cases whether $T_d$ is higher or lower than $M_d$.  Additionally shown, future CMB stage 3 experiments will be sensitive to $\Delta \Neff \sim 0.06$~\cite{Benson:2014qhw,Louis:2016ahn,Suzuki:2015zzg,Grayson:2016smb} and CMB stage 4 experiments hope to reach $\Delta\Neff = 0.027$~\cite{CMBS4}, such that all of our models are discoverable at near-future CMB observatories.  We have assumed that there are no additional degrees of freedom at high energies.

\section{Stochastic Gravitational Waves}

Depending on the scale $V$ and the wall velocities, we can detect stochastic gravitational wave background from the first order phase transition at LISA or future missions BBO or DECIGO (See, {\it e.g.}\/, \cite{Caprini:2015zlo} for a review of the theoretical framework for predictions).

\begin{figure}[t]
\hspace{0.08\textwidth}
\includegraphics[width=0.7\textwidth]{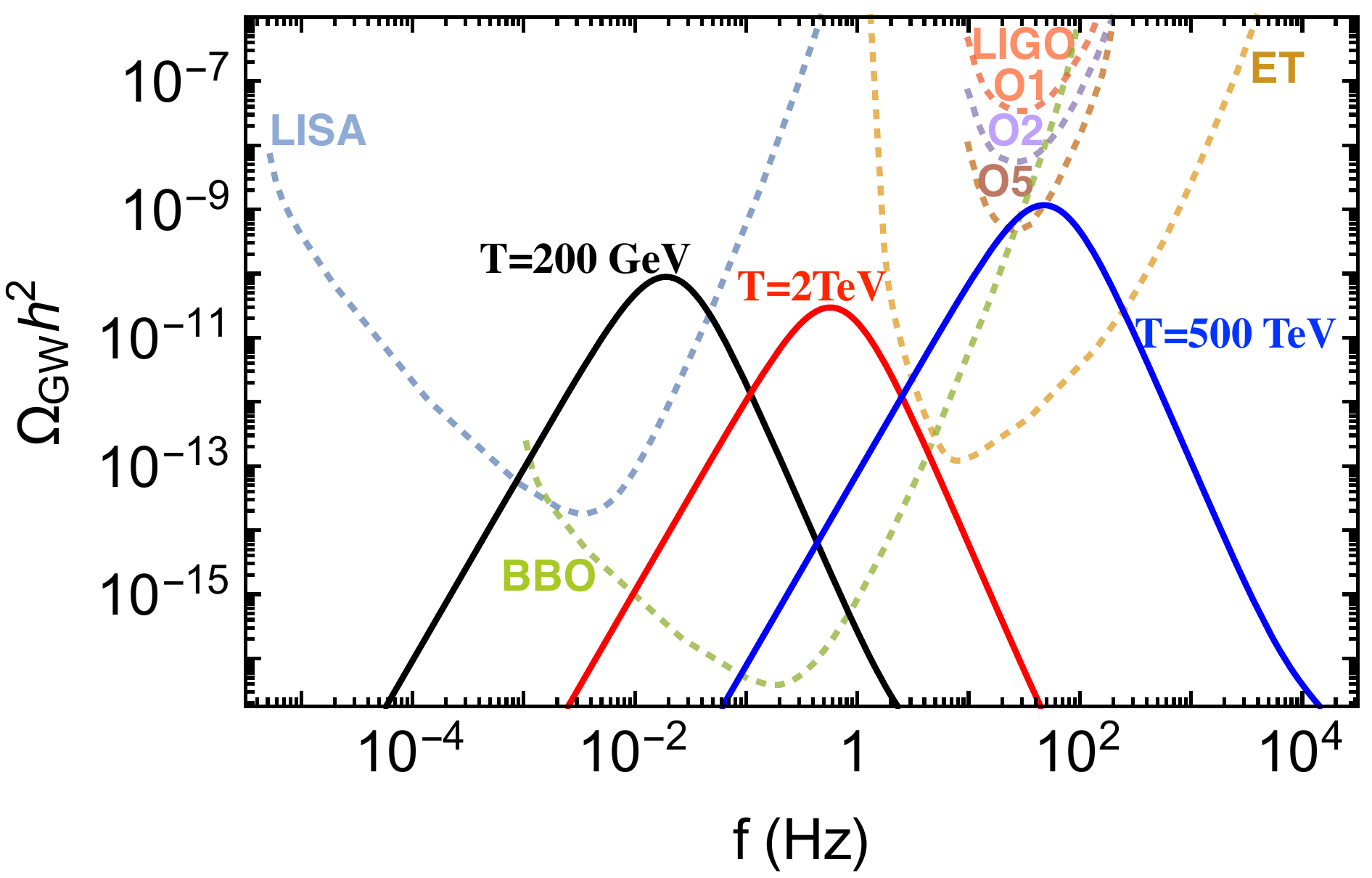}
\caption{GW signal associated with the strong 1st-order phase transitions from the Dark Higgs compared with power law integrated sensitivity curves based on noise curves of LISA~\cite{Audley:2017drz}, LIGO~\cite{Aasi:2013wya}, ET~\cite{Hild:2010id}, and BBO~\cite{Yagi:2011wg}. 
Black: $\alpha=0.5$, $\beta/H=100$.
Red: $\alpha=0.5$, $\beta/H=300$. Blue:  $\alpha=2$, $\beta/H=100$ (see \cite{Caprini:2015zlo} for definitions of $\alpha$ and $\beta/H$). 
In all cases, the bubble wall velocity is $v=0.2$. Such small velocity could occur for a strongly first-order phase transition if there are large friction effects from new degrees of freedom in the plasma with sizeable interactions with the dark Higgs.}
\label{fig:GWsignal}
\end{figure}

Note that the scale of the dark $SU(2)_D$ phase transition $V$ can be much higher than the EW scale without spoiling the baryogenesis.  The peak frequency in the gravitational wave spectrum would be higher in this case, and may be in the Einstein Telescope or even in LIGO/VIRGO/KAGRA windows, see Fig.~\ref{fig:GWsignal}.  For higher scale phase transitions, we may lose collider signatures once $N_d$ is above $m_Z$ and $m_h$ while the $\Neff$ signature remains unchanged.

\section{Conclusion}

We proposed a very simple and minimal model of baryogenesis using a dark $SU(2)_{D}$ gauge group with a first-order phase transition.  Unlike standard EW baryogenesis, it is not subject to the stringent constraints from electric dipole moments.  Yet, it provides verifiable signatures in $\Neff$ at future CMB experiments, as well as  exotic Higgs and $Z$ decays at future $e^{+} e^{-}$ experiments.  Depending on the symmetry breaking scale and the wall velocities, stochastic gravitational waves from the first-order phase transition may be detectable at LISA, or future missions such as ET, BBO and DECIGO.  

\begin{acknowledgments}

We thank Nathaniel Leslie for discussions at the early stages of this work.
This work was supported by the NSF grant PHY-1638509
(H.M.), by the U.S. DOE Contract DE-AC02-05CH11231 (H.M.), by the JSPS
Grant-in-Aid for Scientific Research JP17K05409 (H.M.), MEXT Grant-in-Aid for Scientific Research on Innovative Areas
JP15H05887 (H.M.), JP15K21733 (H.M.), by WPI, MEXT, Japan (H.M.), Hamamatsu Photonics (H.M.), and by the Deutsche Forschungsgemeinschaft under Germany’s Excellence Strategy - EXC 2121 “Quantum Universe” - 390833306. The work of EH was supported by the NSF GRFP.
\end{acknowledgments}

\bibliographystyle{JHEP}
\bibliography{refs}

\end{document}